\documentclass{acm_proc_article-sp}

\newtheorem{lemma}{Lemma}

\def\be{\begin{equation}}
\def\ee{\end{equation}}
\def\bea{\begin{eqnarray}}
\def\eea{\end{eqnarray}}

\def\>{\rangle}
\def\<{\langle}

\def\fp{f_\perp}

\newcommand{\ket}[1]{|#1\>}

\def\<{\langle}
\def\>{\rangle}

\begin{document}

\title{Quantum Digital Signatures}
\numberofauthors{2}

\author{
\alignauthor Daniel Gottesman \\
\email{gottesma@eecs.berkeley.edu\\}	
\affaddr{Computer Science Division \\ University of California \\
		Berkeley, CA 94270}
\alignauthor Isaac L. Chuang
\email{ichuang@media.mit.edu\\}
\affaddr{MIT Media Laboratory \\ 20 Ames Street \\ Cambridge, MA 02139}
}

\maketitle

\begin{abstract}
We present a quantum digital signature scheme whose security is based
on fundamental principles of quantum physics.  It allows a sender
(Alice) to sign a message in such a way that the signature can be
validated by a number of different people, and all will agree either
that the message came from Alice or that it has been tampered with.
To accomplish this task, each recipient of the message must have a
copy of Alice's ``public key,'' which is a set of quantum states whose
exact identity is known only to Alice.  Quantum public keys are more
difficult to deal with than classical public keys: for instance, only
a limited number of copies can be in circulation, or the scheme
becomes insecure.  However, in exchange for this price, we achieve
unconditionally secure digital signatures.  Sending an $m$-bit message
uses up $O(m)$ quantum bits for each recipient of the public
key.  We briefly discuss how to securely distribute quantum public
keys, and show the signature scheme is absolutely secure using one
method of key distribution.  The protocol provides a model for
importing the ideas of classical public key cryptography into the
quantum world.
\end{abstract}

\section{Introduction}

The physics of quantum systems opens a door to tremendously intriguing
possibilities for cryptography, the art and science of communicating in the
presence of adversaries\cite{Bennett87b,Bennett92,Brassard89a,Brassard96a,Bennett98a,Bennett00a,Lo98b,Gottesman2000a}. 
One major goal of classical cryptography is to certify the origin of a
message.  Much like a handwritten signature on a paper document, a
{\em digital signature}\cite{Rivest90,Pfitzmann95a} authenticates an
electronic document and ensures that it has not been tampered with.
The importance of digital signatures to modern electronic commerce has
become such that Rivest has written ``[they] may prove to be one of the most
fundamental and useful inventions of modern cryptography.''\cite{Rivest90}
This is especially true of schemes where the signature can be
recognized using a widely available reference.  The security of all
such public key digital signature schemes presently depends on the
inability of a forger to solve certain difficult mathematical
problems, such as factoring large numbers\cite{Rivest78}.
Unfortunately, with a quantum computer factoring becomes
tractable\cite{Shor97a}, thus allowing signatures to be forged.

We present a {\em quantum} digital signature scheme which is absolutely
secure, even against powerful quantum cheating strategies.  It allows a
sender (Alice) to sign a message
so that the signature can be validated by one or more different
people, and all will agree either that the message came from Alice or
that it has been tampered with.
The scheme described here is somewhat cumbersome, but the underlying
principles suggest novel research directions for the field of quantum
cryptography.  While quantum public keys are more limited than
classical public keys, they remain more powerful than private keys,
and the existence of an unconditionally secure quantum digital
signature scheme suggests an as-yet unrealized potential for quantum
public key cryptography.

Classical digital signature schemes can be created out of any one-way
function\cite{Rompel90a}.  $f(x)$ is a one-way function if it is easy to
compute $f(x)$ given $x$, but computing $x$ given $f(x)$ is very difficult.
This allows the following digital signature scheme\cite{Lamport79a}: Alice
chooses $k_0$ and $k_1$, and publicly announces $f$, $(0,f(k_0))$ and
$(1,f(k_1))$.  Later, to sign a single bit $b$, Alice presents $(b, k_b)$.
The recipient can easily compute $f(k_b)$ and check that it agrees with
Alice's earlier announcement, and since $k_0$ and $k_1$ were known only to
Alice, this certifies that she must have sent the message. 
The public keys can only be used once, unlike more sophisticated
digital signature schemes, but this simple protocol serves as a good
model for a quantum scheme.
While there are many candidate one-way functions, none have been
proven to be secure, and some, such as multiplying together two primes
(the inverse being factoring the product), become insecure on a
quantum computer.  This deficiency leaves a substantial gap in the
cryptographic landscape.

%
%
\section{Main Result}  

We describe a digital signature scheme based on a quantum analogue of
a one-way function which, unlike any classical function, is provably
secure from an information-theoretic standpoint, no matter how
advanced the enemy's computers.  Our goal is to reproduce the primary
advantages of classical public key cryptography in a quantum setting.
These are twofold: First, a public key can be safely given to an
opponent without endangering the security of the protocol.  This
reduces the security requirements involved in key distribution.
Second, every recipient has the same public key, which simplifies key
distribution further; someone who is unsure whether he has received a
correct public key can compare with one from a different source or
with a friend's copy of the key.  Our protocol should not be regarded
as the culmination of this line of research, but as proof of the
principle that quantum protocols can have these properties.

The key idea we introduce is a one-way function whose input is a
classical bit-string $k$, and whose output is a {\em quantum} state
$|f_k\>$ (versus, for instance, a function which maps quantum states
to quantum states).  Like the above classical scheme, we will require
$O(m)$ quantum bits (qubits) to sign a $m$-bit message.  It is not
sufficient, however, to simply plug in $|f_k\>$ in place of $f(k)$.
First, due to the no-cloning theorem\cite{Nielsen00a}, there can be no
perfect equality test for quantum states.  Also, as we show below, the
nature of quantum states provides Alice with non-classical cheating
strategies.  And unlike classical schemes, only a limited number of
copies of the public key can be issued, or the scheme becomes
insecure.  Despite these difficulties, the protocol we present, when
used correctly, allows the probability of any security failure to be
made exponentially small with only polynomial expenditure of
resources.  We begin by defining quantum one-way functions and
discussing their properties.  We then present our signature protocol
and prove its security; this proof appears in two parts, separated by
a discussion of key distribution.  We conclude with some
generalizations of and limitations to our protocol.


%
%
\section{Quantum one-way functions}  

Ever since the invention of secure quantum key
distribution\cite{Bennett92}, many attempts have been made to exploit
the unique properties of quantum systems to provide new cryptographic
primitives.  A great surprise was the failure of quantum bit
commitment\cite{Mayers97a,Lo97a}; subsequently, less powerful but
still interesting primitives such as quantum bit
escrow\cite{Aharonov00a} were introduced.  Looking beyond
cryptography, many more new quantum protocols have been discovered,
such as quantum random access codes\cite{Ambainis99a} and quantum
fingerprints\cite{Buhrman01a}.  

%
%
Here, we introduce a limited-utility quantum one-way function, based
on two properties of quantum systems\cite{Nielsen00a} which are also
essential for quantum fingerprinting.  First, quantum bits, unlike
their classical counterparts, can exist in a superposition of $0$ and
$1$.  The general state of a single qubit is written as a
two-component vector $|\psi\> = \alpha_0 |0\> + \alpha_1 |1\> =
(\alpha_0 \alpha_1)$, where $|0\>$ and $|1\>$ form an orthonormal
basis for the vector space, and $\alpha_0$, $\alpha_1$ are complex
numbers satisfying $|\alpha_0|^2 + |\alpha_1|^2 = 1$.  Because of this
continuous degree of freedom, distances between two qubit states
$|\psi\>$ and $|\psi'\>$ naturally take on non-integer values (less
than the maximum, $1$), defined as $\sqrt{1-|\<\psi|\psi'\>|^2}$,
where $\<\psi|\psi'\>$ is the inner product between the two vectors.
This becomes particularly interesting when considering the general
state of $n$ qubits, $|\psi^n\> = \sum_{j=0}^{2^n-1} \alpha_j |j\>$,
where the number of coefficients is exponentially larger than the
number of qubits.  It follows from simple volumetric arguments that
sets of states $\{|\psi_k^n\>\}$ exist satisfying $|\< \psi_k^n |
\psi_{k'}^n\>| \leq \delta$ for $k\neq k'$ where the set may have many
more than $2^n$ states if $\delta <1$, meaning the states are not
maximally distant from each other.  In fact, as Buhrman, Cleve,
Watrous, and de~Wolf showed\cite{Buhrman01a}, for $\delta \approx
0.9$, one may have a set of size $2^{O(2^n)}$.

%
%
We shall make use of this property by taking all classical bit strings
$k$ of length $L$, and assigning to each one a quantum state $|f_k\>$
of $n$ qubits.  These states are nearly orthogonal: $|\< f_k |
f_{k'}\>| \leq \delta$ for $k \neq k'$, allowing $L$ to be much larger
than $n$.  As mentioned above, for the quantum fingerprint states, $L
= O(2^n)$ with $\delta \approx 0.9$.  Another family is provided by
the set of stabilizer states\cite{Nielsen00a}, with $L = n^2/2 +
o(n^2)$, and $\delta = 1/\sqrt{2}$.  Both these sets are easy to
create with any standard set of universal quantum gates.  A third
family of interest uses just $n=1$ qubit per state, and consists of
the states $\cos (j\theta) |0\> + \sin (j\theta) |1\>$, for $\theta =
\pi/2^L$, and integer $j$.  This family works for any value of $L$,
and gives $\delta = \cos \theta$.

%
%
The second property we exploit is that although the mapping $k \mapsto
|f_k\>$ is easy to compute and verify, it is impossible to invert (without
knowing $k$) by virtue of a fundamental theorem of quantum information
theory.  Holevo's theorem limits the amount of classical information that
can be extracted from a quantum state\cite{Holevo77b,Nielsen00a}; in
particular, measurements on $n$ qubits can give at most $n$ classical bits
of information.  Thus, given $T$ copies of the state $|f_k\>$, we can learn
at most $Tn$ bits of information about $k$, and when $L - Tn \gg 1$, our
chance of successfully guessing the string $k$ remains small.  This means
that $k \mapsto |f_k\>$ acts as a sort of quantum one-way function, with a
classical input and a quantum output.

%
%
Certain important properties of classical functions are taken for
granted which are no longer so straightforward quantum-mechanically.
Given two outputs $|f_k\>$ and $|f_{k'}\>$, how can we be sure that
$k=k'$?  This is done using a simple quantum circuit\cite{Buhrman01a},
which we shall call the {\em swap test}.  Take the states $|f_k\>$ and
$|f_{k'}\>$, and prepare a single ancilla qubit in the state $(|0\> +
|1\>)/\sqrt{2}$.  Next, perform a Fredkin gate (controlled-swap) with
the ancilla qubit as control and $|f_{k'}\>$ and $|f_k\>$ as targets.
Then perform a Hadamard gate on the ancilla qubit and measure it.  If
the result is $\ket{0}$, then the swap test is passed; this always
happens if $|f_{k'}\> = |f_k\>$.  Otherwise, if $|\<f_{k'}|f_k\>|\leq
\delta$, the result $\ket{0}$ occurs with probability at most
$(1+\delta^2)/2$.  If the result is $|1\>$, then the test fails; this
happens only when $k\neq k'$ and occurs with probability
$(1-\delta^2)/2$.  Clearly the swap test works equally well even if
the states are not outputs of the function $f$ --- if the states are
the same, they always pass the swap test, while if they are different,
they sometimes fail.  The point is that an equality test exists, but
fails with nonzero probability.

%
%
Another important property is the ability to verify the output of the
function: given $k$, how do we check that a state $|\psi\> = |f_k\>$?  This
is straightforward: since the function $|k\> |0\> \mapsto |k\> |f_k\>$ is
easy to compute (here, $|0\>$ denotes an $n$ qubit state), simply perform
the inverse operation, and measure the second register.  If $|\psi\> \neq
|f_k\>$, the measurement result will be nonzero with probability $1- |\<
\psi |f_k \>|^2$.  Thus, verification is also possible, but again it is
probabilistic.

%
%
\noindent {\bf A naive quantum signature protocol.}  What happens if
we simply drop in our quantum one-way function in place of the
classical one in Lamport's signature scheme\cite{Lamport79a}
(described above)?  The protocol parameters $L$ and $n$ are fixed, and
a map $k \mapsto |f_k\>$ is chosen by all parties.  Alice generates
$k_0$ and $k_1$ as her private keys, and publicly announces
$(0,|f_{k_0}\>)$ and $(1,|f_{k_1}\>)$ as her public keys.  As in
Lamport's scheme, she then signs a bit $b$ by presenting $(b,k_b)$.
Ideally, the recipient, Bob, would then want to test Alice's quantum
public key for validity.  He can do this using the verification test,
to see if $|k_b\>|f_{f_b}\>$ maps back to $|k_b\>|0\>$.  Furthermore,
once Bob is satisfied with the validity of Alice's message, he would
like to be able to pass it on to Judge Charlie, knowing that Charlie
will also find the message valid.  Unfortunately, Bob's test sometimes
fails; furthermore, it irreversibly consumes one of Alice's public
keys!

Other potential problems arise as well.  For instance, unlike the
output of a perfect classical one-way function, from which someone
with limited computational ability can learn {\em nothing} at all
about the input, $|f_k\>$ always leaks a limited amount of information
about $k$, the input to the quantum one-way function.  Furthermore,
quantum cheating strategies become available; for example Alice may
want to make Bob and Charlie disagree on the validity of her message.
How can we be sure that all of the copies of the public keys she hands
out are identical?  Along the same lines, Alice is free to prepare an
entangled initial state, with which she can delay choosing $k$ until
after she has given $|f_k\>$ away.  Her ability to do this spells the
doom of any attempt to use quantum one-way functions to perform bit
commitment\cite{Lo97a,Mayers97a}, which is one application of
classical one-way functions.  But {\em only} Alice has the ability to
change the state, and it will not help her in this instance.  This
saving grace enables us to use quantum one-way functions to perform
digital signatures.  Most of the new difficulties introduced by
quantum states can be dealt with by using many public keys per message
bit instead of just one.  In the remainder of the paper, we discuss
the details of this modification, and more importantly address the
issue of Alice's quantum cheating strategies.

\section{Quantum signature protocol}

We now present the signature protocol, beginning first with a definition of
what such a protocol should accomplish and how its security is evaluated;
following this we present the quantum protocol in detail.

\noindent {\bf Definition.} We adopt essentially the
usual definition of a one-use digital signature; that is, Alice has a set of
private keys and all recipients have copies of the corresponding public
keys.  Given a message $b$, Alice can then produce a single signed message
$(b,s(b))$.  Conversely, given any message, signature pair $(b',s')$ any
recipient can process the pair to reach one of three possible conclusions:
\begin{quote}
\begin{itemize}
\item[\bf 1-ACC:] Message is valid, can be transferred
\item[\bf 0-ACC:] Message is valid, might not be transferable
\item[\bf REJ:] Message is invalid
\end{itemize}
\end{quote}
The first two results imply that Alice sent the message $b'$.  They
differ in that result {\bf 1-ACC} means the recipient is sure any
other recipient will also conclude the message is valid (thus the
message is ``transferable''), whereas result {\bf 0-ACC} allows the
possibility that a second recipient might conclude the message is
invalid (the number ``1'' or ``0'' refers to the minimum number of
people who agree with the conclusion that the message is valid).  Result {\bf REJ} implies the
recipient cannot safely reach any conclusion about the authenticity of
the message.  We require that any recipient who receives a correct
message, signature pair $(b,s(b))$ always reaches conclusion {\bf
1-ACC}.

\noindent {\bf Security criteria.}  The protocol should satisfy two security
criteria.  First, it should be secure against forging, which means that,
even given access to a valid signed message $(b,s(b))$ and all available
copies of the public keys, no forger has an appreciable chance of creating a
message, signature pair $(b',s')$ (with $b' \neq b$) such that an honest
recipient will accept it (conclusions {\bf 1-ACC} or {\bf 0-ACC}) except
with exponentially small probability.  Second, the scheme should be secure
against Alice's attempts to repudiate the message.  That is, for any pair of
recipients, with high probability, if the first recipient reaches conclusion
{\bf 1-ACC} (the message is valid and transferable), then the second
recipient also reaches conclusion {\bf 1-ACC} or {\bf 0-ACC} (the message is
valid).

Our definition differs from the most common classical definitions in only
three respects: First, the possibility of result {\bf 0-ACC} is not
available in most classical signature schemes (although some allow it).
Second, we only require that the security criteria hold with high
probability (again true of some classical schemes).  Third, and most
notably, the public keys in our scheme are quantum states rather than
classical strings.

%
%
This protocol is applicable to a variety of cryptographic problems.  For
instance, Alice may wish to sign a contract with Bob such that Bob can prove
to Judge Charlie that the contract is valid.  In this case, Bob should
accept the contract whenever he gets result {\bf 1-ACC} for a message, and
Charlie should accept unless he gets result {\bf REJ}.  This problem can
also be solved by a variety of classical protocols, of course.  However,
most only offer security against computationally-bounded attacks.  Others
offer information-theoretic security, but require additional resources
during the key distribution phase (see section~\ref{sec:distribution}), such
as a secure anonymous broadcast channel\cite{Chaum91a} or a noisy
channel\cite{Crepeau90a,Crepeau97a}, which are difficult to justify as physical
resources.  In 
addition, the classical information-theoretic protocols use distinct private
keys and require substantial interaction among the participants during the
key distribution phase, whereas the quantum protocol we present below
requires only a physically plausible quantum channel and modest
interactivity quite similar to that required by classical public key
distribution.


\noindent {\bf Quantum signature protocol specification.} As the
private keys for our protocol, Alice chooses a number of pairs of
$L$-bit strings $\{k_0^i,k_1^i\}$, $1\leq i \leq M$.  The $k_0$'s will
later be used to sign a message $b=0$, and the $k_1$'s will be used to
sign $b=1$.  Note $k_0^i$ and $k_1^i$ are chosen independently and
randomly for each $i$, and $M$ keys are used to sign each bit.  $M$ is
the security parameter; the protocol is exponentially secure in $M$
when the other parameters are fixed.  The states
$\{|f_{k_0^i}\>,|f_{k_1^i}\>\}$ (for each $i$) will then be Alice's
public keys for an appropriate quantum one-way function $f$.  The
public keys are ``public,'' in the sense that no particular security
measures are necessary in distributing them: if a number of copies
fall into the hands of potential forgers, the protocol remains secure.
Note that the creation of these keys is up to Alice (or someone she
trusts), because unknown quantum states cannot be perfectly copied,
according to the no-cloning theorem\cite{Nielsen00a}.  We begin by
making the simplifying assumption that all recipients have received
correct and identical copies of Alice's public keys; we will revisit
this assumption later in the paper.

All participants in the protocol will know how to implement the map $k
\mapsto |f_k\>$.  All participants will also know two numbers, $c_1$ and
$c_2$, thresholds for acceptance and rejection used in the protocol.  A
bound on the allowed value of $c_2$ will be given as part of the proof of
security, below.  $c_1$ can be zero in the absence of noise; the gap
$c_2-c_1$ limits Alice's chance of cheating.  We assume perfect devices and
channels throughout this paper, but our protocol still works in the presence
of weak noise by letting $c_1$ be greater than zero, and with other minor
adjustments.  We further require that Alice limits distribution of the
public keys so that $T < L/n$ copies of each key are available (recall that
$|f_k\>$ is an $n$ qubit state).

Alice can now send a single-bit message $b$ using the following
procedure:

\begin{enumerate}
\vspace*{-0.8ex}

\item Alice sends the signed message $(b,k_b^1,k_b^2,\ldots,k_b^M)$ over an
insecure classical channel.  Thus, Alice reveals the identity of half of her
public keys.
\vspace*{-0.8ex}

\item Each recipient of the signed message checks each of the revealed
public keys to verify that $k_b^i \mapsto |f_{k_b^i}\>$.
Recipient $j$ counts the number of incorrect keys; let this be $s_j$.
\vspace*{-0.8ex}

\item Recipient $j$ accepts the message as valid and transferable (result
{\bf 1-ACC}) if $s_j \leq c_1 M$, and rejects it as invalid (result {\bf
REJ}) if $s_j \geq c_2 M$.  If $c_1 M < s_j < c_2 M$, recipient $j$
concludes the message is valid but not necessarily transferable to other
recipients (result {\bf 0-ACC}).
\vspace*{-0.8ex}

\item Discard all used and unused keys.
\vspace*{-0.8ex}

\end{enumerate}

When $s_j$ is large, the message has been heavily tampered with, and
may be invalid.  When it is small, the message cannot have been
changed very much from what Alice sent.  $s_j$ is similar for all
recipients, but need not be identical.  As we shall see below, the
thresholds $c_1$ and $c_2$ separate values of $s_j$ into different
domains of security.  Forgery is prevented by $c_2$, and cheating by
Alice is prevented by a gap between $c_2$ and $c_1$.


\section{Proof of security I: forgery}  

We need to prove the security of this scheme against two scenarios of
cheaters.  In the first scenario, only Alice is dishonest; her goal is
to get recipients to disagree about whether a message is valid or not
(i.e., she wishes to ``repudiate'' it).
We will show that if one recipient unconditionally accepts ($s_j<c_1 M$),
then it is very unlikely that another will unconditionally reject ($s_{j'}
>c_2M$).  However, we delay this proof until after discussing distribution
of the public keys.

The second scenario is the standard forging scenario.  In this case, Alice and
at least one recipient Bob are honest.  Other recipients or some third party
are dishonest, and they wish to convince Bob that a message $b' \neq b$ is
valid.
Naturally, the forgers can always prevent any message from being received,
or cause Bob to reject a valid message, but we do not consider this to be a
success for the cheaters.

The security proof for this scenario is straightforward.  In the worst
case, the forger Eve has access to all $T$ copies of each public key.
By Holevo's theorem, Eve can acquire at most $Tn$ bits of information
about each bit string $k_b^i$.  When Alice sends the signed message,
Eve may attempt to substitute a different $b' \neq b$ and (possibly)
different values of the $k_{b'}^i$ to go with it.  However, since she
lacks at least $L - Tn$ bits of information about any public key which
Alice hasn't revealed, she will only guess correctly on about $G =
2^{-(L-Tn)} (2M)$ keys.  Furthermore, if she wishes to change a key
whose identity she did not guess correctly, she has only probability
$\delta^2$ of successfully revealing the key.  Each recipient measures
$M$ keys, so when $b \neq b'$, each recipient will find (with high
probability) that at least $(1-\delta^2) (M-G) - O(\sqrt{M})$ public
keys fail.  We will pick $c_2$ so that $(1-\delta^2)(M-G) > c_2 M$,
which means each recipient either receives the correct message, or
rejects the message with high probability.

\section{Key distribution}  
\label{sec:distribution}

For the first scenario, where Alice is dishonest, we will simplify to
the case where there are only two recipients, Bob and Charlie.
However, before tackling the proof, we must return to the issue of key
distribution.  Here, Alice wishes Bob (for instance) to accept the
message and Charlie to reject it.  Certainly, if Alice can give
completely different public keys to Bob and Charlie, she can easily
repudiate her messages; therefore, any signature scheme, classical or
quantum, must be accompanied by a key distribution scheme to eliminate
this possibility.  

Classically, a straightforward assumption is that public keys are
broadcast to all recipients; however, in practice this is seldom the
case (the internet, for instance, is normally used as a point-to-point
network), and creating a cryptographically secure broadcast channel is
a highly nontrivial task.  In the quantum case, we do not even have
the possibility of a broadcast channel, so we must resort to other
means.  One straightforward solution is to assume the existence of a
trusted key distribution center, which has authenticated
links\cite{Barnum01a} to all 
three participants.  Alice sends her public keys to the key
distribution center, which performs swap tests between corresponding
pairs of public keys.  If any pair of public keys fails the swap test,
the center concludes Alice is cheating; otherwise it forwards a copy
of each public key to each recipient.

Alice can prepare any state she wishes for the public keys, including
entangled states and states outside the family $|f_k\>$.  For
instance, she can prepare a symmetric state, such as $\ket{\psi}_B
\ket{\phi}_C + \ket{\phi}_B \ket{\psi}_C$.  Because this state is
invariant under swaps, it always passes all tests, so that the key
distribution center concludes that Bob and Charlie will have the same
key.  But that is an illusion --- clever trickery by Alice who can
nevertheless arrange that Bob and Charlie disagree on the validity of
the corresponding private key $k_{b}^i$.  However, Alice cannot
control which of them receives the valid key; it goes randomly to Bob
or Charlie.  Thus, since $M$ is large, the difference $|s_B - s_C|$ is
$O(\sqrt{M})$ with high probability, which makes it very unlikely that
Bob and Charlie will get definitive but differing results.  That is,
when one of them (say, Bob) accepts a message ({\bf 1-ACC}), that is
$s_B < c_1M$, Charlie almost never rejects it ({\bf REJ}), which would
happen if $s_C > c_2M$.  The gap between $c_1M$ and $c_2M$ protects
them against Alice's machinations.

Of course, assuming a universally trusted third (or in this case,
fourth) party always simplifies cryptographic protocols, so for the
full proof, we wish to consider a more sophisticated scenario.  In
this case, we assume that Bob and Charlie have each received their
public keys directly from Alice, perhaps in person, perhaps via a
private key authenticated channel.  Then to test their keys
classically, Bob and Charlie would announce and compare them.  For our
quantum protocol, they can instead perform the following distributed
swap test:
Each of Bob and Charlie receives from Alice {\em two} copies of
each public key (so there are a total of $T=4$ copies of each public
key in circulation).
For each value of $i$ and $b$, the recipients verify that they all
received the same public key $|f_{k_b^i}\>$.  To do so, each recipient
first performs a swap test between their two keys, then passes one
copy to a single recipient (Bob, for instance).  Bob then checks that
these two test keys pass the swap test as well.  If any keys fail
either test, the protocol is aborted.  Otherwise, discard the test
keys.  The remaining ``kept'' keys are used to verify messages in the main
protocol.

A dishonest recipient can always cause the key distribution phase to
abort, but nothing more.  He could also allow a dishonest Alice to
incorrectly pass the test, but the notion of a digital signature is
only meaningful when at least two participants are honest.  When Alice
is honest, no cheater has an opportunity to alter someone else's
public key, so the scheme remains secure against forgery.

We wish to emphasize that the above suggestions for key distribution
are by no means the only possibilities.  The distributed swap test can
easily be generalized to the case of multiple recipients (see
Section~\ref{sec:generalizations}), for instance, and many classical
methods of key distribution can be adapted for the quantum case to
allow a variety of security assumptions.  Ideally, it would be
possible to state a simple security criterion that would evaluate
whether a given method of key distribution is successful or not
without reference to the particular quantum public key protocol for
which the keys are intended, but we do not attempt to formulate such a
definition.

%
%

\section{Proof of security II: repudiation and transferability}  

We now return to the security of our digital signature protocol, and
show that it prevents Alice from cheating (repudiating a message she
has signed).  Whatever the method of key distribution, some form of
swap test is likely to be present, so we assume the use of a
distributed swap test.  Our goal is to compute the probability $p_{\rm
cheat}$ that Alice can pass all the swap tests but achieve $|s_B -
s_C| > (c_2 - c_1)M$, meaning that Bob and Charlie disagree about the
validity of the message.  We do this by studying a global pure state
$|\Psi\>$, which describes all of the public keys as well as any state
that Alice may have which is entangled with the keys.  Any state which
passes the initial swap tests will be symmetric between the test keys
and the kept keys; in fact, it is symmetric between any individual
test key and the corresponding kept key.  Therefore, we can safely
assume Alice prepares $|\Psi\>$ with this property.  From now on, when
we speak of the swap test, we only refer to the second swap test
between the two test keys.

Now, for each set of four keys (two test and two kept), the most
general state is a superposition of two types of terms.  A type-1 term
may pass the swap test, but leaves Bob and Charlie in agreement, on
average, about the validity of the keys, while a type-2 term
frequently fails the swap test.  To perform the decomposition, we
expand the kept keys and the test keys each in the basis $|f\>|f\>$,
$|\fp^a\>|\fp^{a'}\>$, $|+^a\>$, and $|-^a\>$, where the first ket is
Bob's, the second is Charlie's, $|f\> = |f_{k_b^i}\>$ for the current
value of $b$ and $i$, the states $|\fp^a\>$ form an orthonormal basis
with $|f\>$, and $|\pm^a\> = |f\>|\fp^a\> \pm |\fp^a\>|f\>$.  Thus, a
dishonest Alice might, for instance, prepare the state $|\psi\> =
|f\>_K |f\>_K |+^a\>_T + |+^a\>_K|f\>_T|f\>_T$, where the subscript
$K$ indicates kept keys and $T$ indicates test keys.  A type-1 term
is any term for which both the kept and test keys are in a state
$|f\>|f\>$, $|\fp^a\>|\fp^{a'}\>$, or $|+^a\>$.  Note that a sum of
type-1 terms (such as $|\psi\>$, above) may always pass the swap test,
but also has equal amplitudes for Bob and Charlie to pass key
verification.  A type-2 term is any term including a $|-^a\>$ state
for the kept keys, the test keys, or both.  For the type-2 terms, we
explicitly note the symmetry between the kept and test keys, meaning
the superposition $(|+^a\>_K |-^{a'}\>_T + |-^a\>_K
|+^{a'}\>_T)/\sqrt{2}$ is the only way $|-^a\>|+^{a'}\>$ can appear.
In particular, any sum of type-2 terms respecting this symmetry must
have at least a $50\%$ chance of failing the swap test.  On the other
hand, some superpositions of type-2 terms can give different chances
for Bob and Charlie to pass key verification.  Also note that the
subspace of type-1 terms is orthogonal to the subspace of type-2
terms.

Expanding every set of keys in $|\Psi\>$ in this way gives a global
state which we can again divide up into two terms: $|\Psi_1\> +
|\Psi_2\>$.  Every summand in $|\Psi_1\>$ contains at most $r$ type-2
tensor factors, where $r = cM$ for some constant $c>0$; the rest are
type-1 terms.  $|\Psi_2\>$ consists of terms with more than $r$ type-2
tensor factors.

We wish to show first that for the state $|\Psi_1\>$, $|s_B-s_C|$ will
be small.  If there were only a single summand in $|\Psi_1\>$, this
would clearly be true, since each type-1 factor has an equal
probability of contributing to $s_B$ and to $s_C$, and there are only
a few type-2 terms.  However, different summands of $|\Psi_1\>$ with
different patterns of type-1 and type-2 states might interfere
quantum-mechanically.  To show $|s_B - s_C|$ is small even in this
case, it will be sufficient to look just at the kept keys, and
furthermore only at the states $|+^a\>$ and $|-^a\>$ --- the states
$|f\>|f\>$ and $|\fp^a\>|\fp^{a'}\>$ always produce exactly the same
contribution to $s_B$ and to $s_C$, and cannot interfere with each
other.  In fact, $|+^a\>$ and $|-^{a'}\>$ cannot interfere when $a
\neq a'$, so we can restrict attention to just two states $|+\>$ and
$|-\>$.  On the other hand, the combination $|0\> = (|+\> +
|-\>)/\sqrt{2}$ will always cause Bob's key to pass and Charlie's to
fail, whereas $|1\> = (|+\> - |-\>)/\sqrt{2}$ gives the opposite
result.  

This allows us to simplify the problem.  We can invoke the lemma from
Appendix~\ref{sec:lemma} to show that, for large $M$ and sufficiently
small $c$, the probability that $|s_B - s_C| < (c_2 - c_1)M$ for
$|\Psi_1\>$ is exponentially small in $M$; less than $2^{-[1 -
H((1-c_2+c_1)/2) - H(c)]M}$, in fact, although this is not at all a
tight bound.

For $|\Psi_2\>$, we wish to show that the probability of passing the
swap test is very small.  To see this, it will suffice to consider a
modified swap test which passes any state of the test keys except
$|-^a\>$; certainly the probability of passing this test can be no
smaller than the probability of passing the original swap test.  Since
each type-2 term by itself has at least a $1/2$ chance of failing the
modified swap test, a tensor product of $r$ or more of them passes
with probability no larger than $2^{-r}$.  There can be no
interference between different positions for the type-2 terms, since
the modified swap test is compatible with the projection onto type-1
and type-2 terms.  Therefore, the probability of $|\Psi_2\>$ passing
the swap test (original or modified) is at most $2^{-r} = 2^{-cM}$.
Since $c>0$, this is exponentially small in $M$.

Now we can put this together to obtain a bound on $p_{\rm cheat}$,
which is the probability that the state both passes the swap test and
produces $|s_B - s_C| > (c_2 - c_1)M$.  The $|\Psi_1\>$ term might
have a good chance of passing all swap tests, but yields an
exponentially small chance of giving the required separation between
$s_B$ and $s_C$.  The $|\Psi_2\>$ term might have $O(1)$ probability
of having $|s_B - s_C| > (c_2 - c_1)M$, but only has an $O(2^{-r})$
chance of passing all swap tests.  The best case for constructive
interference between the two terms gives a total probability $p_{\rm
cheat}$ at most twice the sum of the two probabilities for $|\Psi_1\>$
and $|\Psi_2\>$, which is still exponentially small in $M$.
Therefore, Alice has $p_{\rm cheat} \sim O(d^{-M})$ probability of
successfully cheating for some $d>1$.


\section{Generalizations and extensions}  
\label{sec:generalizations}

One straightforward generalization is to use the distributed swap test
with many recipients.  To do this, we can replace the swap test with a
test for complete symmetry of $s$ states~\cite{Buhrman01a}.  Instead
of preparing an ancilla in the state $(|0\> + |1\>)/\sqrt{2}$, we
prepare a superposition over states indexed by all permutations of $s$
elements.  Then perform the permutation $\sigma$ conditioned on the
ancilla being in the state $|\sigma\>$, and finally measure the
ancilla to see if it remains in the original superposition.  If the
state of keys being tested is completely symmetric, it always will
pass, otherwise it has some chance to fail.  Furthermore, note that,
for any state of the keys, the probability that any particular pair of
keys out of the $s$ keys being tested will fail a regular swap test is
no larger than the chance that the full set of keys fails the symmetry
test, so for any pair of keys, the symmetry test is at least as
sensitive as the swap test.

The distributed symmetry test then allows public key distribution for
$t>2$ recipients.  Each person receives $t+1$ copies of each public
key (so there are $T = t(t+1)$ copies in circulation) and tests them
for complete symmetry.  Assuming they pass, each recipient keeps one
copy of each key to verify a signature, sends one copy to each of the
other recipients to perform a second symmetry test, and keeps the last
copy for his own symmetry test.  Each recipient now has $t$ test keys,
and performs a symmetry test on those keys.  He rejects the set of
keys if it fails either of the symmetry tests he performed.  If we
restrict attention to any particular pair of recipients, this
procedure essentially reduces to the distributed swap test again, so
the proof of the previous section tells us that for any two recipients
$i$ and $j$, the probability that the keys pass the symmetry test but
$|s_i - s_j| > (c_2 - c_1)M$ is exponentially small in $M$.  This
shows the signature protocol remains secure with the distributed
symmetry test and many recipients.

We can also create additional thresholds to allow more than one
transfer.  That is, $0=c_0 < c_1 < \ldots < c_q < 1$, and if $c_{r-1}
< s_i < c_{r}$, for $r \leq q$, then recipient $j$ will {\em
$(q-r)$-accept} the message.  When a recipient $s$-accepts the message
(result {\bf $s$-ACC}), he is convinced the message is valid (it
originated with Alice), and that any other recipient will at least
$(s-1)$-accept the same message.  A recipient who $0$-accepts a
message is convinced it is valid, but is not sure someone else will
agree with him (result {\bf 0-ACC}).  In other words, $s$-acceptance
means the recipient is sure he can convince $s$ other people of the
message's validity sequentially, even if each wants to be sure later
people accept it as well.  The security of $s$-acceptance follows
immediately from the proof of security in the last section, simply
substituting $c_r - c_{r-1}$ for $c_2 - c_1$.

Another useful extension is the ability to expand the original
symmetry test to additional groups of keys.  Assume we have a single
recipient Bob who communicates with two separate sets of recipients;
we will assume Bob is not allied with the sender Alice.  Suppose Bob
receives $s+1$ keys originally, and performs a test on them for
complete symmetry.  Then he keeps one key for verifying a signature,
and uses some (but not all) of the $s$ test keys to perform a
distributed symmetry test with a group $R_1$ of recipients.  The extra
test keys certainly do not affect the security of the distributed
symmetry test.  Suppose Bob uses the remaining test keys to perform a
distributed symmetry test with a second group of recipients $R_2$,
possibly at a much later time.  Then if Charlie is in $R_1$ and Diane
is in $R_2$, we know that (with high probability) either the keys fail
a test, or that $|s_B - s_C|<\Delta M$ and that $|s_B - s_D| < \Delta
M$ for some $\Delta$, in which case it is also true that $|s_C - s_D|
< 2 \Delta M$.  In other words, even though Charlie and Diane have not
interacted directly, the gap between $s_C$ and $s_D$ is still bounded,
but by twice the margin between two recipients in the same group.

While we have described a procedure for signing single-bit messages,
multi-bit messages can be sent by repeating the process, using $M$
pairs of public keys for each message bit.  However, a much more
efficient procedure is to first encode the message in a classical
error-correcting code with distance $M$, and to use a single pair of
public keys for each encoded bit.  The single-bit protocol can be
viewed as a special case of this using a repetition code.  Valid
messages are codewords of the error-correcting code; to change from
one valid message to another requires altering $M$ bits.  Therefore,
the above security proofs hold with only two changes: $G$, the number
of keys successfully guessed by Eve, is now $2^{-(L-Tn)} (2N)$, where
$N$ is the length of the full encoded message.  In addition, if Alice
attempts to cheat, she can produce a difference $|s_B - s_C|$
proportional to $N$, not $M$, using type-1 terms.  We should thus have
$M$ scale linearly with $N$ when the latter is very large.


\section{Conclusions}  

The digital signature scheme provided here has many potential
applications.  It combines unconditional security with the flexibility
of a public key system.  An exchange of digital signature public keys
is sufficient to provide authentication information for a quantum key
distribution session.  Quantum digital signatures could be used to
sign contracts or other legal documents.  In addition, digital
signatures are useful components of other more complex cryptographic
procedures.

One particularly interesting application is to create a kind of
quantum public key cryptography.  If Bob has Alice's public key, but
Alice has nothing from Bob, then Bob can initiate a quantum key
distribution session with Alice.  Bob will be sure that he is really
talking to Alice, even though Alice has no way to be sure that Bob is
who he says he is.  Therefore, the key generated this way can be
safely used to send messages from Bob to Alice, but not vice-versa.

However, quantum public keys have a number of disadvantages.  It is
not possible to sign a general unknown quantum state, even with
computational security~\cite{Barnum01a}; this is unfortunate, since a common
classical method for distributing public keys is to have a trusted
``Certificate Authority'' (whose public key is already well-known)
sign them for later transmissions.  However, perhaps this can be
circumvented: the quantum public keys of our protocol are {\em known}
quantum states, so perhaps there is some way to securely sign them for
distribution at a later time.

Note that in a purely classical scheme, the public key can be given
out indiscriminately.  This cannot be true of a quantum scheme: when
there are very many copies of a public key, sufficiently careful
measurements can completely determine its state, and therefore one may
as well treat the public key as classical.  In that case, security
must be dependent on computational or similar assumptions.  Thus, any
quantum digital signature scheme will necessarily require limited
circulation of the public key.  This is primarily a question of
efficiency, since sufficiently large $L$ allows many keys to be
issued.

So how do the required resources scale with the number of recipients?
There are three resources that one might consider: the size of each
public key, the size of a single signed message, and size of the
private key.  In our case, a single public key need only scale as the
logarithm of the number of receivers.  This is good, since the public
keys are made of qubits, which may be the most expensive component.
However, the length $L$ of the private key must be at least equal to
$T$, the total number of public keys in circulation, which must be
linear or quadratic in the number of recipients.  It remains possible
that an improved proof or protocol could reduce the required $L$
substantially, although we are not optimistic on this point.  However,
this is not too serious, since the classical memory used to store the
private key is already quite cheap.  A more serious flaw in our
current protocol is the requirement that the length of a signed
message also scale linearly with $L$.  There does not seem to be a
fundamental requirement for this, luckily, so it seems plausible that
an improved protocol is possible for which the length of messages
scales at most logarithmically with $T$.

Scaling of resources with other variables can probably be improved as
well.  Earlier, we showed how to reduce the amount of key required to
send long messages; perhaps further improvement is possible.
Classical private-key authentication allows the expenditure of only a
logarithmic amount of key in the length of the message; it is
reasonable to speculate that similar efficiency could be achieved
here.  Efficiently signing known quantum states would allow this, for
instance, since then we could sign a quantum
fingerprint~\cite{Buhrman01a} of the message.

Since our scheme requires a new set of keys for each message, the
total amount of key consumed also scales linearly with the number of
messages sent.  It would be preferable to reduce this to the levels
allowed by classical protocols: the log of the number of messages, or
even constant (although it seems unlikely that is possible).  Either
would imply substantial reuse of public keys.  Designing such a
protocol will be a difficult task, however, since usual classical
techniques for reusing signature keys cannot be applied to quantum
public keys.

In summary, we have demonstrated the existence of an unconditionally
secure public key digital signature scheme, something which is not
possible classically.  Many potential improvements remain, however.
The possibilities and ultimate limitations of quantum public key
cryptography remain largely unexplored.

\noindent {\bf Acknowledgements:} We thank C. Bennett, C. Crepeau,
D. DiVincenzo, D. Leung, H. K. Lo, D. Mayers, M. Mosca, J. Smolin,
B. Terhal, and W. van Dam for helpful comments.  DG was supported by a Clay
long-term CMI prize fellowship.  ILC was supported in part by the Things
That Think consortium and the QuARC project under a DARPA Quantum
Information Science and Technology grant.


\appendix

\section{Lemma for proof of transferability}
\label{sec:lemma}

\begin{lemma}
For any $\Delta > 0$, there exists a $c>0$ such that, for large $M$,
the following holds: Given a state $|\Psi_1\>$ of $M$ qubits which is
a sum of tensor products of $|+\>$ and $|-\>$ with at most $r=cM$
$|-\>$ factors in any term, then measurement in the $|0\>$, $|1\>$
basis will with high probability produce a result with weight between
$M(1/2 - \Delta)$ and $M(1/2 + \Delta)$.
\end{lemma}

That is, if we have a superposition of words of weight at most
$r$, the weight measured in the Hadamard-rotated basis will be near
$M/2$.

{\bf Proof}: This is easy to show.  There can be at most about $N =
{M \choose r}$ terms in the sum $|\Psi_1\>$.  Note that $\log_2 N
\approx M H(r/M) = M H(c)$, where $H(x)$ is the Hamming entropy $H(x)
= -x \log_2 x - (1-x) \log (1-x)$.  Thus, the probability
$|\<y|\Psi_1\>|^2$ to get a particular string $y$ in the $|0\>$,
$|1\>$ basis is at most $N/2^M \approx 2^{(H(c) - 1)M}$.  Since
there are about $2 {M \choose M(1/2 - \Delta)}$ strings outside the
allowed range, the total probability of being outside the allowed
range is at most about
\begin{equation}
2^{1 + [H (1/2 - \Delta) + H(c) - 1] M}.
\end{equation}
This is small for large $M$ whenever
\begin{equation}
H(1/2 - \Delta) + H(c) < 1.
\end{equation}
Since $H(1/2 - \Delta) < 1$ for any $\Delta > 0$, and $H(c)
\rightarrow 0$ as $c \rightarrow 0$, the lemma follows.

\end{document}